\documentclass[%
 reprint,
 amsmath,amssymb,
 aps,
 pra,
showkeys,
showpacs
]{revtex4-1}

\usepackage[dvipsnames]{xcolor}
\usepackage[normalem, normalbf]{ulem}
\usepackage{graphicx}
\usepackage{dcolumn}
\usepackage{bm}
\usepackage{amsmath, amsthm,amssymb}
\usepackage{changes}
\usepackage{epstopdf}

\begin{document}

\title{Towards thermal noise free optomechanics}

\author{Michael A. Page}

\email{michael.page@uwa.edu.au}

\address{University of Western Australia, 35 Stirling Highway, Crawley, Western Australia, 6009, Australia}

\author{Chunnong Zhao, David G. Blair, Li Ju}

\address{University of Western Australia, 35 Stirling Highway, Crawley, Western Australia, 6009, Australia}

\author{Yiqiu Ma}

\address{California Institute of Technology, 1200 E California Blvd, Pasadena, CA 91125, United States}

\author{Huang-Wei Pan, Shiuh Chao}

\address{National Tsing Hua University, No. 101, Section 2, Guangfu Road, East District, Hsinchu City, Taiwan, 300, Republic of China}

\author{Valery P. Mitrofanov}

\address{Faculty of Physics, Lomonosov Moscow State University, Moscow, 119991, Russia}

\author{Hamed Sadeghian}

\address{The Netherlands Organization for Applied Scientific Research, TNO, Stieltjesweg 1, 2628 CK, Delft, The Netherlands}

\vspace{10pt}

\begin{abstract}

Thermal noise generally greatly exceeds quantum noise in optomechanical devices unless the mechanical frequency is very high or the thermodynamic temperature is very low. This paper addresses the design concept for a novel optomechanical device capable of ultrahigh quality factors in the audio frequency band with negligible thermal noise. The proposed system consists of a minimally supported millimeter scale pendulum mounted in a Double End-Mirror Sloshing (DEMS) cavity that is topologically equivalent to a Membrane-in-the-Middle (MIM) cavity. The radiation pressure inside the high-finesse cavity allows for high optical stiffness, cancellation of terms which lead to unwanted negative damping and suppression of quantum radiation pressure noise. We solve for the optical spring dynamics of the system using the Hamiltonian, find the noise spectral density and show that stable optical trapping is possible. We also assess various loss mechanisms, one of the most important being the acceleration loss due to the optical spring. We show that practical devices, starting from a centre-of-mass pendulum frequency of 0.1 Hz, could achieve a maximum quality factor of \(10^{14}\) with optical spring stiffened frequency 1-10 kHz. Small resonators of mass 1 \(\mu\)g or less could achieve a Q-factor of \(10^{11}\) at a frequency of 100 kHz. Applications for such devices include white light cavities for improvement of gravitational wave detectors, or sensors able to operate near the quantum limit.

\end{abstract}


%
\vspace{2pc}

%
%
%
%

\maketitle
\section{Introduction}\label{Introduction}

Optomechanics, the interaction of photons with phonons, is a phenomenon that allows for sensitive measurements close to fundamental limits \cite{AspelmeyerReview2014}.One example of extremely precise measurement using optomechanics is the detection of gravitational waves in large interferometers \cite{LSCLIGO, Aasi2013}. Optomechanics can also be used to impose and control quantum states on high frequency resonators at cryogenic temperatures \cite{OConnellQGS, TeufelQGS, ChanQGS}. Achievement of quantum measurement and control in low frequency regimes and higher temperatures, however, has been limited by quantum and thermal noise \cite{BraginskySQL}. This paper argues that it is possible to overcome these sources of noise using a device with a resonance which lies in the audio band, and without having to apply extreme cryogenic cooling. This is achieved through enhancement of the quality factor using optical spring stiffness to store energy of vibration in a lossless field.

The use of optical forces as restoring forces of mechanical resonators has been investigated in various configurations from 1g mirror pendulums stiffened with a linear optical spring \cite{CorbittDilution}, picogram mirror resonators suspended from fibres \cite{NiTrap}, to optically levitated sub-micron particles \cite{ChangTrapping, LiNanosphere} and mirrors \cite{LamTripod}. However, systems using linear optical spring effects will ultimately be limited by quantum radiation pressure noise induced heating. The sub-micron levitated particle could be an ideal quantum measurement system itself, but is not suitable for amplification of a Gaussian beam signal. The trapped picogram mirror is limited by modes other than the centre-of-mass (CM) pendulum mode. The acceleration loss due to the coupling between the optical spring and the mirror internal modes will be a limiting factor for all systems. These limiting factors are addressed in this paper.

\begin{figure*}
\begin{center}
\includegraphics[width = 0.9\textwidth]{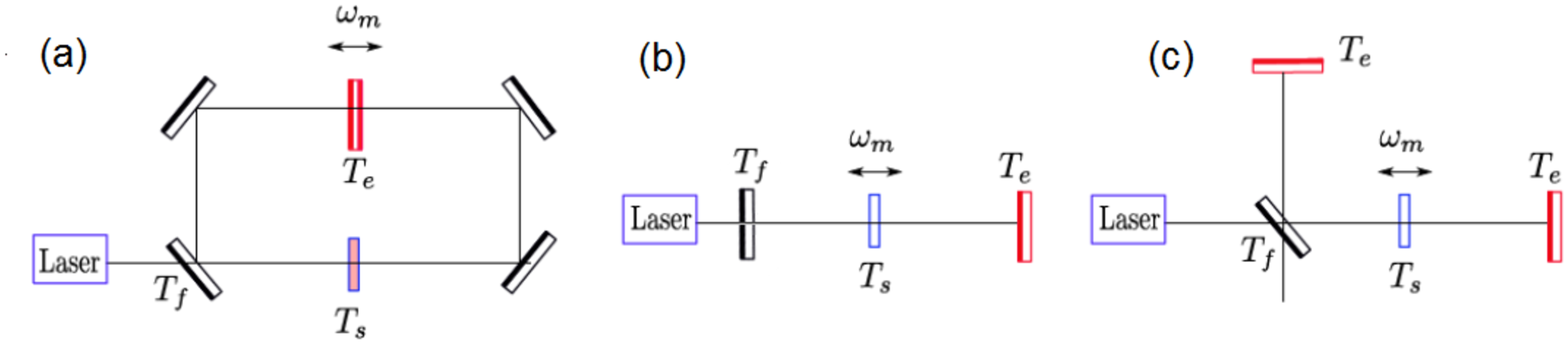}
\end{center}
\caption{\textbf{(a)} Double End-Mirror Sloshing (DEMS) cavity. The cavity is separated into two subcavities by the mechanical resonator of natural frequency \(\omega_m\) and sloshing mirror of transmissivity \(T_s\). Light enters through the input mirror with transmissivity \(T_f\). The mechanical resonator, referred to as the end mirror, has transmissivity \(T_e << T_f\) \textbf{(b)} The MIM configuration in which a low mass, high \(Q_m\) transmissive membrane creates a coupled cavity. \textbf{(c)} Coupled cavity which can be used to model the quantum dynamics of both \textbf{(a)} and \textbf{(b)}.\label{Trapping}}
\end{figure*}

Aspelmeyer \textit{et al.} \cite{AspelmeyerReview2014} pointed out that the minimum requirement for quantum optomechanics at room temperature is that the product of Q-factor \(\times\) frequency exceeds \(6 \times 10^{12}\). The principle of optical dilution involves using the radiation pressure from a laser to apply an optical spring stiffness to a mechanical resonator. Since the optical field is lossless, the stiffness and mechanical frequency of the resonator are increased without adding any extra sources of mechanical loss. The increase in quality factor due to the optical spring is given by the ratio of the optical spring stiffness to the original mechanical stiffness.

One of the most exciting applications of the technology proposed here is to create mechanical resonators of sufficiently low noise that they can be used to realise white light cavities for resonant enhancement of gravitational wave detectors. The white light cavity allows resonant amplification of a broad band of frequencies through the use of optomechanically induced negative dispersion. This has been shown by Miao \textit{et al.} \cite{HaixingFilters} to allow a substantial improvement in sensitivity of gravitational wave detectors between 100 Hz and 1 kHz. However, this form of resonant amplification of the signal is only useful if the thermal noise of the mechanical resonator can be effectively eliminated.

The above considerations define the basic requirements for practical thermal noise free optomechanical devices. The device size should be considerably larger than the laser wavelength to allow light processing without large wavefront distortion. The resonator should be mechanically suspended to eliminate the need to control multiple degrees of freedom, but the suspension should have extremely low mass to minimise the thermal noise contribution from the thermal reservoir. The internal acoustic modes of the device should be high compared to the optical spring frequency to prevent internal acoustic loss contributions. The device should be able to withstand the optical power required to achieve a high optical spring frequency. Once thermal noise is eliminated, quantum radiation pressure noise becomes an issue. The design presented here also has the advantage of minimising quantum noise while simultaneously cancelling terms that lead to optical spring induced instabilities. This paper addresses the optimisation of a resonator capable of meeting all of the above requirements.

\begin{figure}
\begin{center}
\includegraphics[width = 0.4\textwidth]{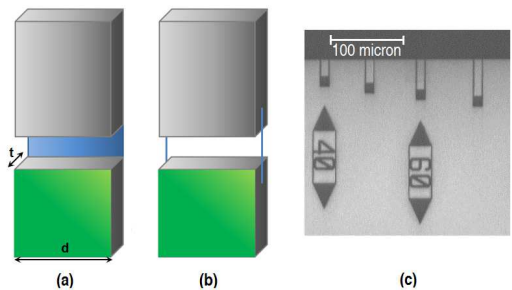}
\end{center}
\caption{A concept of the resonator to be used in this experiment. Grey pieces are silicon, blue are suspension materials and green are high reflectivity coatings \textbf{(a)} Original ``cat-flap'' concept of a pendulum swinging from a thin silicon nitride membrane. In this paper, we use a square face for simplicity of description and presentation of estimates, characterised by side length \textit{d}. The aspect ratio is the ratio of the thickness \textit{t} to the square face size \textit{d}. \textbf{(b)} Alternate suspension with bonded carbon nanotubes. \textbf{(c)} Micrograph of resonators suspended by silicon nitride nanowires, fabricated at Van Leeuwenhoek Laboratory in Delft \cite{HamedMicrofab}. Note that the resonator size is approximately 10 \(\mu\)m, which would cause diffractive losses if used in an optical cavity. \label{cflap}}
\end{figure}

\begin{figure}
\begin{center}
\includegraphics[width = 0.4\textwidth]{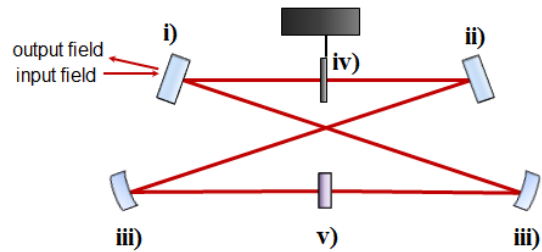}
\end{center}
\caption{Practical implementation of the DEMS cavity. \textbf{(i):} Flat input mirror, transmissivity \(T_f\). \textbf{(ii):} Flat mirror, transmissivity \(T_m\). \textbf{(iii):} Curved mirror, transmissivity \(T_n\). \textbf{(iv):} Mechanical resonator, transmissivity \(T_e\). \textbf{(v):} Sloshing mirror, transmissivity \(T_s\). Achieving the desired quantum dynamics requires that \(T_f >> T_s >> T_e\). Keeping \(T_m\) and \(T_n\) as low as possible is also desirable for obtaining high finesse. \label{Bowtie}}
\end{figure}

We present and analyse the design of a novel device incorporated in a Double End-Mirror Sloshing (DEMS) cavity shown in Figure \ref{Trapping}. This device consists of a pendulum where the suspended mass made of some material with low acoustic loss such as silicon. The suspension can be carbon nanotubes, silicon nitride nanowires or membranes. Section \ref{Theory} describes the quantum optomechanics for one of these devices mounted in a coupled cavity. In the same way that a membrane-in-the-middle (MIM) cavity couples two optical modes to a single mechanical resonator (the membrane), the DEMS cavity replaces the membrane with a high reflectivity mirror. The original concept of our resonator is the ``cat-flap'' resonator shown in Figure \ref{cflap}(a), and has since been extended to a resonator suspended by nanowires, shown in Figure \ref{cflap}(b) and (c). The cavity coupling is adjustable using a separate sloshing mirror of finite transmission, which also comprises the end mirrors of the pair of Fabry-Perot cavities created in this way.  Diagonal mirrors are used for input-output coupling. For practical construction the DEMS cavity can be configured in a bowtie structure shown in Figure \ref{Bowtie}. 

Our DEMS cavity and the MIM can be modelled with a configuration that is topologically equivalent to both. The arrangement of the DEMS cavity is shown to result in extra input noise channels compared to the MIM cavity, but the optical spring mechanics remain the same. The DEMS cavity allows us to experiment upon a macroscopic block with nanoscale suspension, a novel area of research, unlike the MIM cavity which requires an extremely thin, semi-transparent mechanical resonator.

Sections \ref{AccelLoss} and \ref{Comparisons} present an analysis of two important loss mechanisms of the resonator. The strong optical spring and high frequency motion of the pendulum cause acceleration losses that increase as the optical spring frequency approaches internal modes of the suspended mass. This loss occurs because the radiation pressure force acting on the pendulum mass causes compression of the resonator material, due to its own inertia and elastic compressibility. Tensile stresses in the suspension result in violin string modes that will also contribute to the mechanical loss. Both of these will place an upper limit upon the achievable maximum quality factor. We show the acceleration loss effect on the optically diluted Q-factor of resonators with masses ranging from 1 \(\mu\)g to 2 mg.

Section \ref{Thermoelastic} outlines thermoelastic losses for resonators constructed with various materials. Thermoelastic damping occurs for flexural members. One face will bend concave, causing compression on the surface, while the opposite face will bend convex, causing stretching of the atomic lattice. The localised change in volume causes heat flow, which has the consequence of mechanical loss \cite{ZenerThermo, ZenerThermo2, LifshitzThermo}. We show the thermoelastic loss limit of resonators constructed from crystalline and amorphous forms of silicon, silicon dioxide and silicon nitride, and give strategies to reduce this limit.

Section \ref{SuspensionFabrication} gives schemes for the construction of silicon nitride membrane and nanowire suspensions. Techniques used in the construction include chemical vapour deposition of silicon nitride, lithography, and wet and dry etching.

\section{Quantum dynamics of the DEMS cavity}\label{Theory}

Ma, \textit{et al.} \cite{YiqiuDilution} showed that a coupled membrane-in-the-middle (MIM) cavity is capable of producing a thermal and quantum noise-free optical trap, while maintaining stability through reducing unwanted optical spring anti-damping. The analysis for this paper refers to the DEMS cavity, shown in Figure \ref{Trapping} (a), and its equivalent, the MIM cavity, shown in Figure \ref{Trapping} (b). The equivalent optomechanical cavity used for analysis is shown in Figure \ref{Trapping} (c).


\subsection{Optical spring dynamics of a coupled cavity}\label{Quantum}

The Hamiltonian of the cavity shown in Figure \ref{Trapping} (c) can be written in terms of the cavity mode operators, mechanical operators and input noise terms:

\begin{multline}\label{Hamiltonian}
\hat{H} = \hbar\omega_c(\hat{a}^{\dag}\hat{a} + \hat{b}^{\dag}\hat{b}) + \hbar\omega_s(\hat{a}^{\dag}\hat{b} + \hat{a}\hat{b}^{\dag}) + \frac{\hat{p}^2}{2m}+\frac{1}{2}m\omega_m^2\hat{x}^2 \\
+ \hbar G_0 \hat{x} (\hat{a}^{\dag}\hat{a} + \hat{b}^{\dag}\hat{b}) + H^{opt}_{ext} + H_{ext}^m
\end{multline}

Here, \(\hat{a}\) and \(\hat{b}\) are the annihilation operators for the cavity modes in the left and right sub-cavity. The first two terms in the Hamiltonian describe photons circulating in these subcavities with resonant frequency \(\omega_c\). The third term, a cross term of \(\hat{a}\) and \(\hat{b}\), describes photons transferring from one subcavity to the other at the sloshing frequency \(\omega_s = c\sqrt{T_s}/L \). \(\hat{x}, \hat{p}\) are the position and momentum operators of the vibrating mirror, respectively. Optomechanical coupling is represented by the term featuring the product of \(\hat{x}\) with the cavity modes \(\hat{a}\) and \(\hat{b}\). The coupling coefficient \(G_0 = \omega_0/L\) links the optical and mechanical operators, with \(\omega_0\) being the laser frequency. The second last term is light input:

\begin{align}\label{Hext}
\hat{H}^{opt}_{ext} &= i\hbar\sqrt{2\gamma_f}(\hat{a}^{\dag}(\hat{a}_{in,1} + \hat{a}_{in,2}) - h.c.) \nonumber \\
&+\hbar\sqrt{2\gamma_e}(\hat{b}^{\dag}(\hat{b}_{in,1} + \hat{b}_{in,2}) - h.c.)
\end{align}

and \(H^m_{ext}\) corresponds to the coupling of the mechanical resonator to the environment. Here, \(\gamma_f = cT_f/2L_c\) and \(\gamma_e = cT_e/2L_c\) are the transmissive losses through the front (input) and end mirrors, with \(T_f\) and \(T_e\) being the respective transmissivities. The end mirror in Figure \ref{Trapping} (c) is equivalent to the optomechanical mirror in the DEMS cavity.

The \(H^{opt}_{ext}\) term differs between a DEMS and MIM cavity. The input noise terms \(\hat{a}_{in}\) and \(\hat{b}_{in}\) are split into two independent input channels in the DEMS cavity due to the input laser entering the front mirror at an angle. Ideally, the mirrors \textbf{ii} and \textbf{iii} shown in Figure \ref{Bowtie} should have \(T_m, T_n << T_f\), with the cavity only pumped through one mirror, so the \(\hat{b}_{in}\) terms can be set to zero for simplicity.

Solving the dynamics of the above system gives a frequency dependent optical spring stiffness \(K_{opt}(\omega)\) \cite{YiqiuDilution}. This can be expanded in a series of \(\omega\):

\begin{equation}\label{ComplexSpring}
K_{opt}(\omega) \simeq -K_{opt}(0) - \frac{\partial K_{opt}(\omega)}{\partial \omega}\omega - \frac{1}{2}\frac{\partial^2 K_{opt}(\omega)}{\partial \omega^2}\omega^2
\end{equation}

where the first term on the right hand side is used to derive the optical trapping frequency \(\omega_{opt}\), which is the mechanical resonance modified by the optical spring stiffness:

\begin{equation}\label{OSFreq}
\omega^2_{opt} = \frac{\hbar G_0^2 \bar{a}^2_t}{m\omega_s\gamma_f} + o(\epsilon)
\end{equation}

\(\bar{a}_t\) is related to the trapping beam power \(P_{trap}\) through \(\bar{a}_t = \sqrt{P_{trap}/\hbar\omega_0}\). The \(o(\epsilon)\) here describes all higher order terms with \(\epsilon \sim \frac{\Delta_t - \omega_s}{\omega_s}, \gamma_e/\gamma_f\), with \(\Delta_t\) being the cavity detuning.

This optical spring effectively increases the stiffness without introducing losses. The resonator Q is proportional to the ratio of total energy stored to energy lost per cycle. With extra stiffness provided by the optical spring, the system’s stored energy consists of two parts, the energy stored in the optical field (proportional to the optical spring constant) and the mechanical energy (proportional to the mechanical spring constant). The energy lost per cycle is proportional to the mechanical energy only. The quality factor increases as:

\begin{align}\label{Dilution}
Q_{opt} &= Q_{m}\frac{K_{opt}}{K_{m}} \nonumber \\
&= Q_{m}\left(\frac{\omega_{opt}}{\omega_{m}}\right)^2
\end{align}

where \(Q_m\), \(K_m\) and \(\omega_m\) are the mechanical Q-factor, mechanical stiffness and zero-gravity resonant frequency of the pendulum, respectively, with no applied laser. \(Q_{opt}\) is the Q-factor obtained after applying the optical spring and \(K_{opt} = m\omega_{opt}^2\) is the optical spring constant. This equation gives the impression that we can raise the optical stiffness as high as allowed by power constraints. In practice, the nonuniform radiation pressure force that creates the optical spring will excite other mechanical modes, such as the mirror internal modes and suspension mechanical modes, etc. These mechanical modes are associated with unavoidable losses that contribute to the optical spring loss through radiation pressure force and hence the limit of the final quality factor to \(Q_f\).
The effect of the mechanical modes on the diluted Q-factor are analysed in Sections \ref{AccelLoss} and \ref{Comparisons}.

The optical spring is not purely a linear spring. The Taylor expansion of the spring constant \(K_{opt}(\omega)\) in the frequency domain contains imaginary part and higher order terms. The imaginary part of the spring constant contributes to the optical damping that can be used to cool Brownian motion, but it can also cause instability if the imaginary part of the spring constant is negative and not sufficiently balanced by the mechanical damping.

The optical (anti)-damping rate \(\Gamma\) is:

\begin{equation}\label{Damping}
\Gamma = \frac{16\hbar G_0^2 \bar{a}^2_t}{m \gamma_f^2 \omega_s} \times \frac{\Delta_t - \omega_s}{\omega_s} - \frac{8\hbar G_0^2 \bar{a}^2_t }{m \omega_s^2}\times \frac{\gamma_e}{\gamma_f} + o(\epsilon^2)
\end{equation}

When \(\Delta_t = \omega_s\) and \(\gamma_e = 0\), the optical damping is mostly cancelled. We can still choose system parameters to preserve stability should the end mirror not be perfectly reflective.

The effect of the second order term of Taylor expansion of \(K_{opt}(\omega)\) to \(\omega\) is similar to mechanical inertia, or so called optomechanical inertia \cite{Farid}. The main contribution is at the zeroth order of \(\epsilon\):

\begin{equation}\label{Inertia}
m_{opt} = \frac{\hbar G_0^2 \bar{a}_t^2}{\gamma_f\omega_s^3} + o(\epsilon)
\end{equation}

\(m_{opt}\) acts as a negative inertia, and thus the effect must be kept small in order to prevent instability.

Some parameters for the DEMS cavity are shown in Table \ref{mass}, including those relative for calculating optomechanical inertia. For a microgram pendulum with a trapping power of 1.6 W, corresponding to an optical spring frequency \(\omega_{opt} = 2\pi \times 100\) kHz, the negative inertia is seen to be approximately 1\% of the resonator's mass and thus will not cause instability.

\begin{table}\begin{center}\begin{ruledtabular}
\begin{tabular}{|c|c|c|}
\textbf{Parameter}&\textbf{Symbol}&\textbf{Value} \\ \hline
resonator mass & m & 2 \(\mu\)g \\ \hline
cavity half length & L & 500 mm \\ \hline
input mirror transmissivity & \(T_f\) & \(10^4\) ppm \\ \hline
sloshing mirror transmissivity & \(T_s\) & 100 ppm \\ \hline
end mirror transmissivity & \(T_e\) & 1 ppm \\ \hline
intracavity power & \(P_{trap}\) & \(1.6\) W \\ \hline
mechanical Q factor & \(Q_m\) & \(10^6\) \\ \hline
mechanical resonant frequency & \(\omega_m\) & \(2\pi\times 0.1 \) Hz \\ \hline
environmental temperature & \(T_{env}\) & 300 K \\ \hline
\end{tabular}\end{ruledtabular}\end{center}
\caption{Sample parameters for a trapped resonator. }
\label{mass}
\end{table}

\begin{table}\begin{center}\begin{ruledtabular}
\begin{tabular}{|c|c|c|}
\textbf{Parameter}&\textbf{Symbol}&\textbf{Value} \\ \hline
cavity bandwidth & \(\gamma_f\) & \(2\pi \times 0.5\) MHz \\ \hline
sloshing frequency & \(\omega_s\) & \(2\pi \times 1\) MHz \\ \hline
optical spring frequency & \(\omega_{opt}\) & \(2\pi \times 100\) kHz \\ \hline
final Q factor & \(Q_{f}\) & \(10^{11}\) \\ \hline
optomechanical inertia & \(m_{opt}\) & \(1.3\times 10^{-2}\) \(\mu\)g \\ \hline
\end{tabular}\end{ruledtabular}\end{center}
\caption{Some derived parameters for the trapped resonator using the values from Table \ref{mass}. The final quality factor is the Q-factor enhanced by optical dilution but limited by acceleration losses described in Section \ref{AccelLoss}}
\label{Param2}
\end{table}

\begin{figure}\begin{center}
\includegraphics[width = 0.4\textwidth]{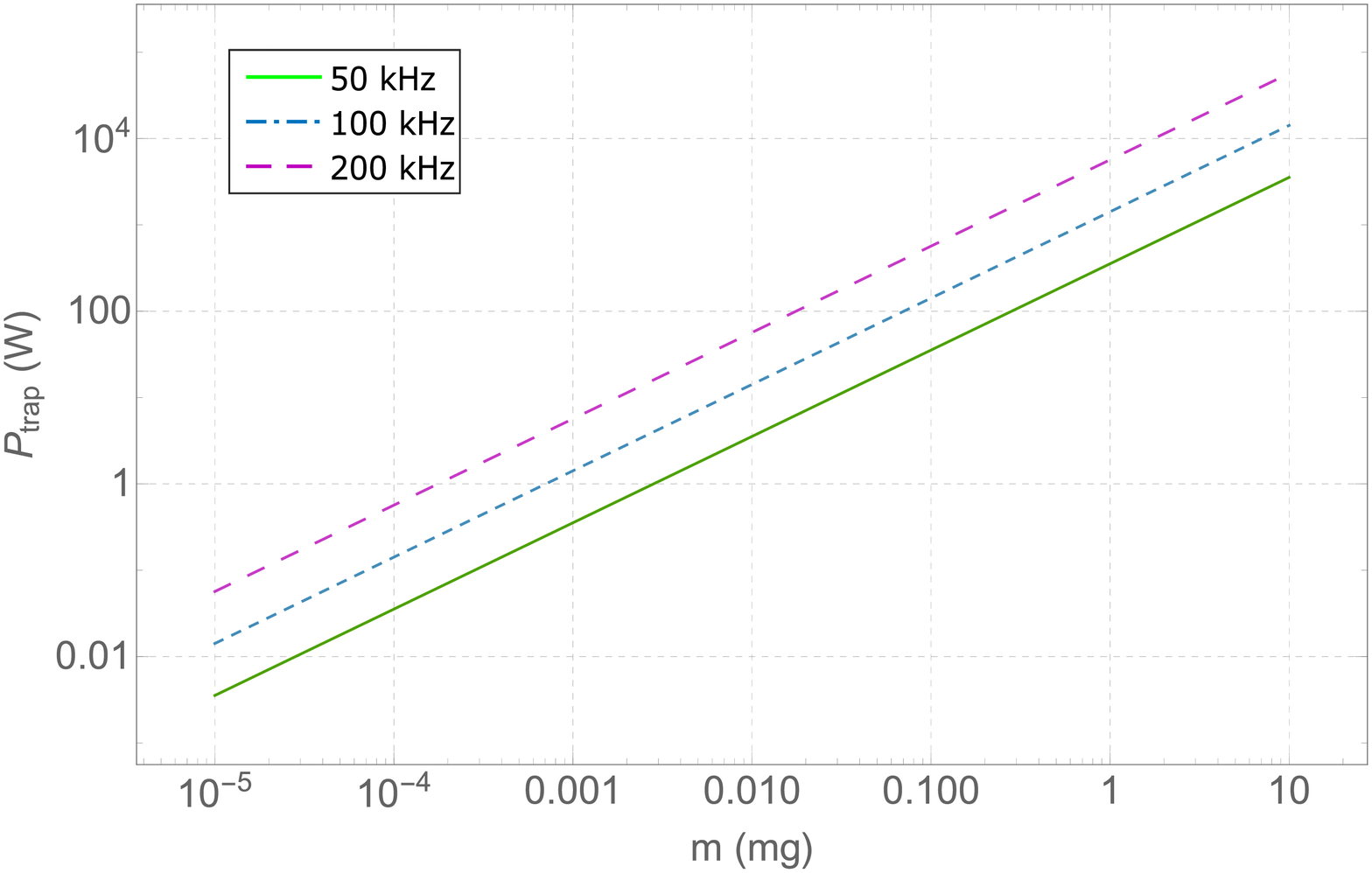}
\end{center}
\caption{Using Equation \ref{OSFreq} and Table \ref{mass} the circulating power required to reach the given optical spring frequencies is plotted against the mass of the resonator. \label{opticMassPower}}
\end{figure}

\subsection{Quantum radiation pressure cancellation}\label{QRPN}

Quantum radiation pressure noise (QRPN) is another issue associated with optical dilution. It will drive the resonator randomly, which is equivalent to increased environmental temperature. The quantum radiation pressure noise spectrum is given by \cite{YiqiuDilution}:

\begin{equation}\label{BA}
S_{FF}^{rad} = \frac{2\hbar G_0^2 \bar{a}_t^2}{\omega_s^2}\frac{\gamma_e}{\gamma_f} + \frac{8 \hbar^2 G_0^2}{\gamma_f \omega_s^4}(\Delta_t^2 - \omega_s^2)^2 + o(\epsilon^2)
\end{equation}

Where the subscript FF denotes a double-sided force noise spectrum. To cancel the quantum noise, we require the end mirror transmissivity to be as low as possible compared to the input mirror transmissivity. Typically the transmissivity of the input mirror can be \(10^4\) ppm, while it is possible to achieve 10 ppm transmissivity for the end mirror using conventional Nd:YAG coatings, and even less than 10 ppm with state-of-the-art coatings \cite{ColeCrystalline2}. This will reduce the ratio \(\gamma_e/\gamma_f\) and allow for cancellation of the first term. The second term can be cancelled by choosing an appropriate detuning.

This effect can also be seen in \(\hat{a}^m_{out}\), the part of the outgoing field which contains the displacement information, under the assumption that the end mirror is perfectly reflective:

\begin{equation}\label{Displace}
\hat{a}^m_{out} = -2i G_0 \bar{a}_{in}\hat{x} + 2i\frac{G_0 \bar{a}_{in}\omega_s^2}{\Delta_t^2}\hat{x}
\end{equation}

The first term on the right hand side is the field reflected directly from the trapped mirror while the second is the field transmitted out of the cavity. Thus, the output field is not disturbed by the displacement of the resonator when the cavity detuning is equal to the sloshing frequency.



\section{Effects of Mechanical losses}\label{MechN}
\subsection{Acceleration loss}\label{AccelLoss}

Thus far we have shown how an optical spring can be used to dilute thermal noise. However, the coupling of the optical spring to mirror internal modes, and other mechanical modes associated with suspension mediated by nonuniform radiation pressure force will introduce losses to the system.
These losses will place a limit upon the maximum achievable Q-factor. We will use a simple 1-dimensional analysis to show how the final Q-factor is related to internal mode frequencies. Summing the loss factors of material loss and attachment loss, we obtain the total loss \(Q^{-1}_{tot}\):

\begin{equation}\label{QPendulum}
Q^{-1}_{tot} = \frac{\Delta W_{mass}+\Delta W_{sus}+\Delta W_{bond}}{2\pi(W_{mech} + W_{opt})}
\end{equation}

where \(W_{mech}\) is the elastic energy stored in the mass and suspension, \(W_{opt}\) is the energy stored in the optical spring, \(\Delta W_{mass}, \Delta W_{sus}\) are the energies lost by the mass and suspension per cycle and \(\Delta W_{bond}\) is the energy lost per cycle by the connection of the suspension to the support. For order of magnitude estimation we neglect optical spring loss and torsional motion, and ignore the energy stored in the gravitational field which is negligible in the case of optical spring damping. Assuming operation in the regime \(W_{mech} << W_{opt}\), the stored mechanical energy becomes negligible and \(Q^{-1}_{tot}\) becomes:

\begin{equation}\label{QPendulum2}
Q^{-1}_{tot} = \frac{\Delta W_{mass}+\Delta W_{sus}+\Delta W_{bond}}{2\pi W_{opt}}
\end{equation}

When radiation pressure accelerates the resonator, deformation propagates throughout the mass. The associated acceleration loss \(Q^{-1}_{acc}\)
leads to a reduction in the final Q-factor. A part of the energy associated with the optical spring transfers to the mechanical resonator through the optical spring restoring force \(F_{opt} = K_{opt}z\) and is stored in the mechanical resonator. We now estimate \(Q^{-1}_{acc}\) as follows. The energy stored in the optical spring is:

\begin{equation}\label{OpticEnergy}
W_{opt} = \frac{1}{2}K_{opt}z_0^2 = \frac{1}{2}M\omega_{opt}^2 z_0^2
\end{equation}

where \(M = \rho A t\) is the mass, with density \(\rho\), \(A = d^2\) is the surface area of the square mirror with side length \(d\), t is the thickness of the mirror and \(z_0\) is the oscillation amplitude.

 By considering only the mirror's longitudinal internal modes, we can estimate the change of the resonator's elastic energy \(\delta W_{e}\) when the centre of mass moves to the maximal deflection \(z_0\). Assuming for a thick resonator in which the energy is proportional to uniform 1D compressive stress \(K_{opt}z_0/A\), the change in mechanical energy becomes:

\begin{equation}\label{deltaW}
\delta W_{e} = \frac{B A t}{2Y}(\frac{K_{opt}z_0}{A})^2
\end{equation}

where \((\frac{K_{opt}z_0}{A})^2/2Y\) is the elastic energy density of a uniformly compressed beam, Y is Young's Modulus of the silicon mirror, and B is a constant of proportionality. The energy lost per period of oscillation is \(\delta W_{e}\) multiplied by the loss angle \(\phi\) of the mass's material:

\begin{equation}\label{BigDeltaW}
\Delta W_{mass} = \phi \frac{B t}{2 Y A}(K_{opt} z_0)^2
\end{equation}

Using \ref{BigDeltaW} and \ref{OpticEnergy}, the loss factor of the resonator can be written as:

\begin{equation}\label{Loss}
Q_{acc}^{-1} = \frac{\phi B \rho t^2 \omega_{opt}^2}{Y}
\end{equation}

The lowest longitudinal mode of a plate of thickness t depends on the thickness, density and Young’s Modulus as follows:

\begin{equation}\label{longModes}
\omega_{int} = \frac{\pi}{t}\sqrt{\frac{Y}{\rho}}
\end{equation}

Introducing this into Equation \ref{Loss}, the loss factor becomes:

\begin{equation}\label{DilutionLimit}
Q^{-1}_{acc} \sim \phi B (\frac{\omega_{opt}}{\omega_{int}})^2
\end{equation}

We can simplify the term \(\phi B\) as the intrinsic loss of the resonator's internal modes \(Q_{mirror}\). The upper limit of this term can be taken as the acoustic loss of the high reflectivity coating, which is \(\sim 10^{-4}\) for conventional SiO\(_2\)/Ta\(_2\)O\(_5\) coatings and \(\sim 10^{-6}\) for crystalline coatings \cite{ColeCrystalline}. Thus, we have the relation:

\begin{equation}\label{DilutionLimit2}
Q^{-1}_{acc} \sim Q^{-1}_{mirror} (\frac{\omega_{opt}}{\omega_{int}})^2
\end{equation}

 \begin{figure}\begin{center}
\includegraphics[width = 0.4\textwidth]{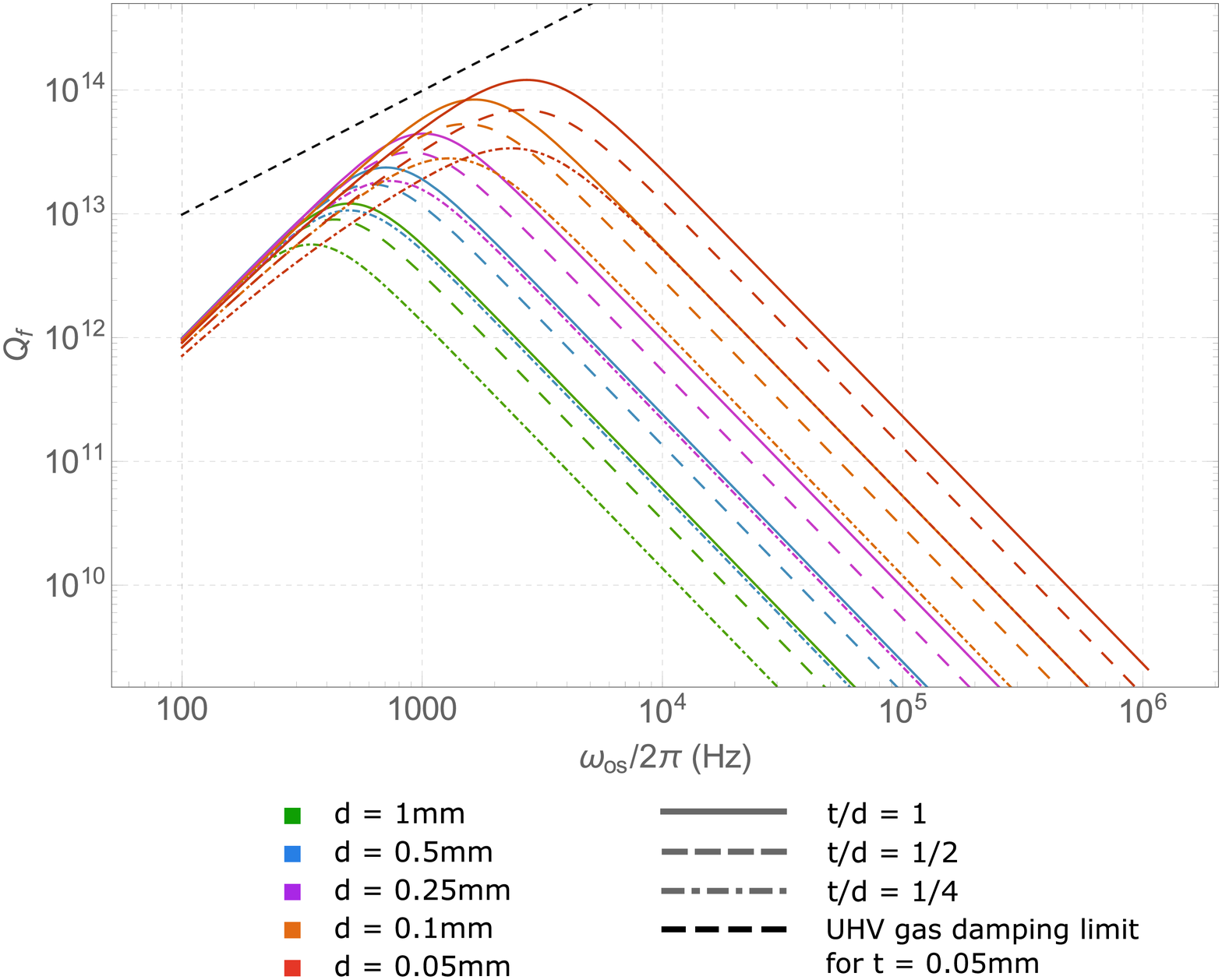}
\caption{Acceleration loss limited quality factor \(Q_f\) of the mechanical resonator vs angular frequency at UHV pressure of \(10^{-11}\) Torr. The dashed black line shows \(Q_{air}\) of a 50 micron thick resonator at the same pressure. Quality factor of the resonator increases with the optical spring effect and decreases with acceleration loss according to Equation \ref{DilutionLimit2}. The graphs above are for silicon mirrors suspended by SiN nanowires, with square mirror side length \textit{d} and aspect ratio \textit{t/d} as shown in Figure \ref{cflap}. The starting Q-factor \(Q_m\) is \(10^6\) for SiN at room temperature \cite{ZwicklSiN}, and the acceleration losses are inversely proportional to internal mode \(Q_{mirror} = 10^6\). \label{PendulumGraphs}}
\end{center}
\end{figure}

Equation \ref{DilutionLimit2} sets a limit to the achievable Q-factor determined by the compressibility and losses of the mirror, which we describe as the acceleration loss. This limit results in some optimal \(\omega_{opt}\). The internal frequencies of the resonator will depend on the size, controlled by \(d\) and the aspect ratio \(t/d\). We have modelled the internal frequencies of resonators with dimension \(d\) from 0.05 mm to 1 mm, and \(t/d\) of 1, 1/2 and 1/4. The zero-gravity resonance frequency of a pendulum, \(\omega_m\), is found through the following equation \cite{RoarksStress}:

\begin{equation}\label{ZeroGravity}
\omega_m \simeq 1.732\sqrt{\frac{Y I}{m L_c^3}}
\end{equation}

which is an approximate equation for the fundamental frequency of a cantilever with an end mass \(m\), length \(L_c\), second moment of area \(I\), provided that it has uniform linear mass density. The quality factor of the cantilever mode increases as per Equation \ref{Dilution}. As the optical spring frequency increases, the effects of acceleration loss increase, and the acceleration loss-limited Q-factor is:

\begin{equation}\label{FinalQ1}
\frac{1}{Q_{acc,lim}} = \frac{1}{Q_{opt}}+\frac{1}{Q_{acc}}
\end{equation}

The \(Q_{acc,lim}\) for a SiN nanowire suspended resonator can theoretically reach higher than \(10^{14}\) for a 50 micron cubed mirror with low-loss crystalline mirror coatings. However, gas damping from the residual air pressure limits the maximum possible Q-factor even at experimentally feasible Ultra High Vacuum (UHV), so an extra term must be added to obtain the final quality factor \(Q_f\):

\begin{equation}\label{FinalQ}
\frac{1}{Q_f} = \frac{1}{Q_{opt}}+\frac{1}{Q_{acc}}+\frac{1}{Q_{air}}
\end{equation}

The gas damping induced Q-factor limit \(Q_{air}\) of a vibrating cantilever is \cite{PressureQ}:

\begin{equation}
Q_{air} = \frac{\omega_{opt} \eta}{4 P_{air}}\sqrt{\frac{\pi R_{gas} T_{env}}{2 M_{air}}}
\end{equation}

Where \(P_{air}\) is the pressure surrounding the resonator, \(\eta\) is the mass per unit area of the face with the normal vector in the direction of oscillation, \(R_{gas}\) is the universal gas constant and \(M_{air}\) is the molar mass of air. For a nanowire suspended cat-flap resonator, the inertia of the bulky mirror means that \(\eta = \rho \times t\) can be used. For a pressure of \(10^{-11}\) Torr, which has been used in atom trapping experiments \cite{UHV1}, the \(Q_{air}\) for a 50 micron cubed mirror is approximately \(5\times 10^{14}\) at 10 kHz and \(5 \times 10^{15}\) at 100 kHz. This is satisfactory compared to the \(Q_{acc,lim}\) of \(1\times 10^{14}\) and \(3\times 10^{11}\) for a 50 micron cube resonator at 10 kHz and 100 kHz respectively. Thicker resonators will have a lower acceleration loss Q-factor limit and higher gas damping Q-factor limit. At 100 kHz, the \(Q_{air}\) limit at a more easily achievable pressure of \(10^{-8}\) Torr is \(5\times 10^{12}\) for a 50 micron cubed mirror and \(1\times 10^{13}\) for a 100 micron cubed mirror.

An estimation of the acceleration loss limited quality factor obtainable from optical trapping and ultra high vacuum is shown in Figure \ref{PendulumGraphs}. With silicon nitride nanowire suspension the limit of the quality factor due to acceleration loss and UHV pressure is of the order \(Q_f = 10^{14}\) for resonators with size \(d\) of 100 microns for frequencies 1-10 kHz. A resonator with \(d < 0.1\) mm is required in order to achieve a maximum Q-factor of approximately \(3 \times 10^{11}\) at 100 kHz. While decreasing the size increases the maximum \(Q_{acc,lim}\), smaller resonators have increased thermoelastic and gas damping loss due to smaller thickness, more suspension loss due to lower suspension mode frequencies, and require a smaller beam spot to minimise diffraction loss.

\subsection{Suspension loss}\label{Comparisons}

\begin{figure}\begin{center}
\includegraphics[width = 0.4\textwidth]{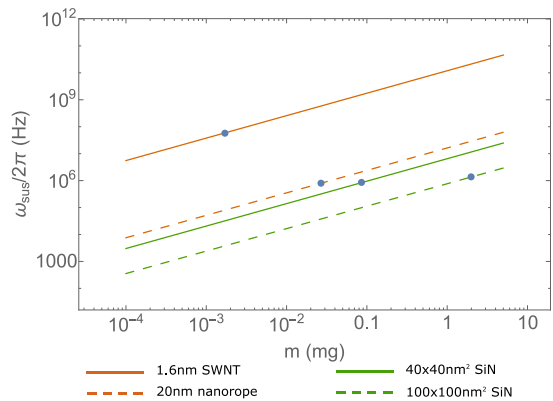}
\caption{\(n=1\) violin string frequencies of various suspension types. The blue dots indicate the mass which induces 20\% of the breaking stress of the material. The highest frequencies belong to the single walled nanotube of diameter 1.6 nm and thickness 0.34 nm. The 20 nm carbon nanorope is composed of approximately 90 SWNTs \cite{YuNanotubes}. The silicon nitride nanowires are labelled according to their cross-sectional dimensions. The high breaking stress of SiN LPCVD fabrication is used. As seen in Figure \ref{PendulumGraphs}, non-CM mode frequencies must be kept above 1 MHz, and preferably above 10 MHz, in order to achieve suitable optical dilution at frequencies of the order of 100 kHz. Single wall carbon nanotubes are preferable for keeping the internal mode frequencies as high as possible, but if the coupling of violin string modes to the CM motion is low, it will be more beneficial to choose highly stressed SiN nanowires for their higher material Q-factor. \label{SuspensionGraphs}}
\end{center}\end{figure}

The suspension is another source of mechanical loss. It transfers mechanical energy from the pendulum centre-of-mass motion to other modes. To simulate suspension losses, we calculate the length of a cantilever beam, with a mass on the free end, that has a fundamental frequency of 0.1 Hz, which is chosen to ensure a large dilution factor \(\omega_{opt}/\omega_m\). This fundamental mode is modified by the optical spring effect, and its frequency is shifted close to the optical spring frequency. The suspension internal modes are then found from tensed string mechanics with the well known equation \(\omega_n = (n\pi/L_z)\sqrt{T/\mu}\), where T is the tension, which can be found from the mass of the resonator, \(L_z\) is the length of suspension and \(\mu\) is the linear mass density. These modes are also referred to as ``violin-string'' modes.

Four different suspension types are analysed throughout this section. The first type of suspension is single walled carbon nanotubes (SWNTs). They have an extremely high breaking stress, exceeding 20 GPa \cite{LiNanotubes}, but their small diameter of 1-2 nm makes them extremely weak. The second type is carbon nanoropes constructed from several SWNTs, whose properties have been studied by Yu, \textit{et al.} \cite{YuNanotubes}. Due to intermolecular forces, the entire rope can be considered as a filled cylinder for the purposes of calculating moment of inertia and linear mass density. However, most of the load is applied to the perimeter nanotubes, making the strength per unit area less than for a single nanotube. The minimum breaking strength of nanoropes is 11 GPa \cite{YuNanotubes}. The number of nanotubes in total and on the perimeter can be calculated using a prescription given by Yu, \textit{et al.} \cite{YuNanotubes}. The next two types of suspension are silicon nitride nanowires of differing dimensions. The nanowires should be fabricated by Low Pressure Chemical Vapour Deposition (LPCVD) in order to obtain the highest possible breaking stress of 5-8 GPa \cite{Chao9, Chao10}.

The \(n=1\) violin string mode frequencies for nanowire and nanotube suspension are shown in Figure \ref{SuspensionGraphs}. SWNTs give the highest violin string frequency at a given mass, but due to their extremely small size (typical SWNT diameter is 1.6 nm) are very weak compared to other suspension types. Nanoropes have much lower violin string frequencies due to their inefficiency in transferring loads.

The suspension types shown here can be designed for fundamental violin string frequencies of greater than 1 MHZ. The losses into the suspension can be reduced by centre of percussion tuning of the optical spring as demonstrated by Braginsky \textit{et al.} in 1998 \cite{BraginskyPercussion}. They demonstrated that a reduction factor of 100 could be achieved. We note also that because the suspension has very high tension, the acoustic modes have Q-factors higher than the material Q-factor \cite{MitrofanovDilution, CagnoliDilution}. The physics of suspension losses for the cat-flap resonator is very similar to that of losses and noise due to the fibre suspensions used for test masses in gravitational wave detectors.

\subsection{Thermoelastic and other losses}\label{Thermoelastic}

\begin{figure}\begin{center}
\includegraphics[width = 0.4\textwidth]{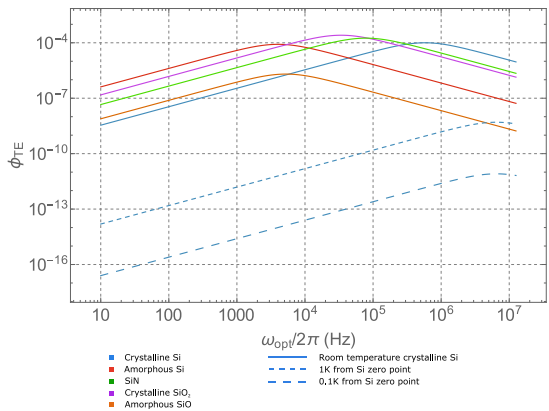}
\end{center}
\caption{Thermoelastic loss curves for a 0.1 mm cube mirror constructed of various materials. Loss is plotted against optical spring frequency. Crystalline silicon has the best room temperature thermoelastic loss properties at up to 6 kHz for this particular resonator, and \(\alpha\)-SiO\(_2\) is the best for frequencies above 6 kHz. The thermodynamic values around the zero point of coefficient of thermal expansion for crystalline Si are derived from Okada \textit{et al.} \cite{OkadaSiliconExpansion} and Glassbrenner \textit{et al.} \cite{GlassbrennerSilicon}. \label{Materials}}
\end{figure}

\begin{figure}\begin{center}
\includegraphics[width = 0.4\textwidth]{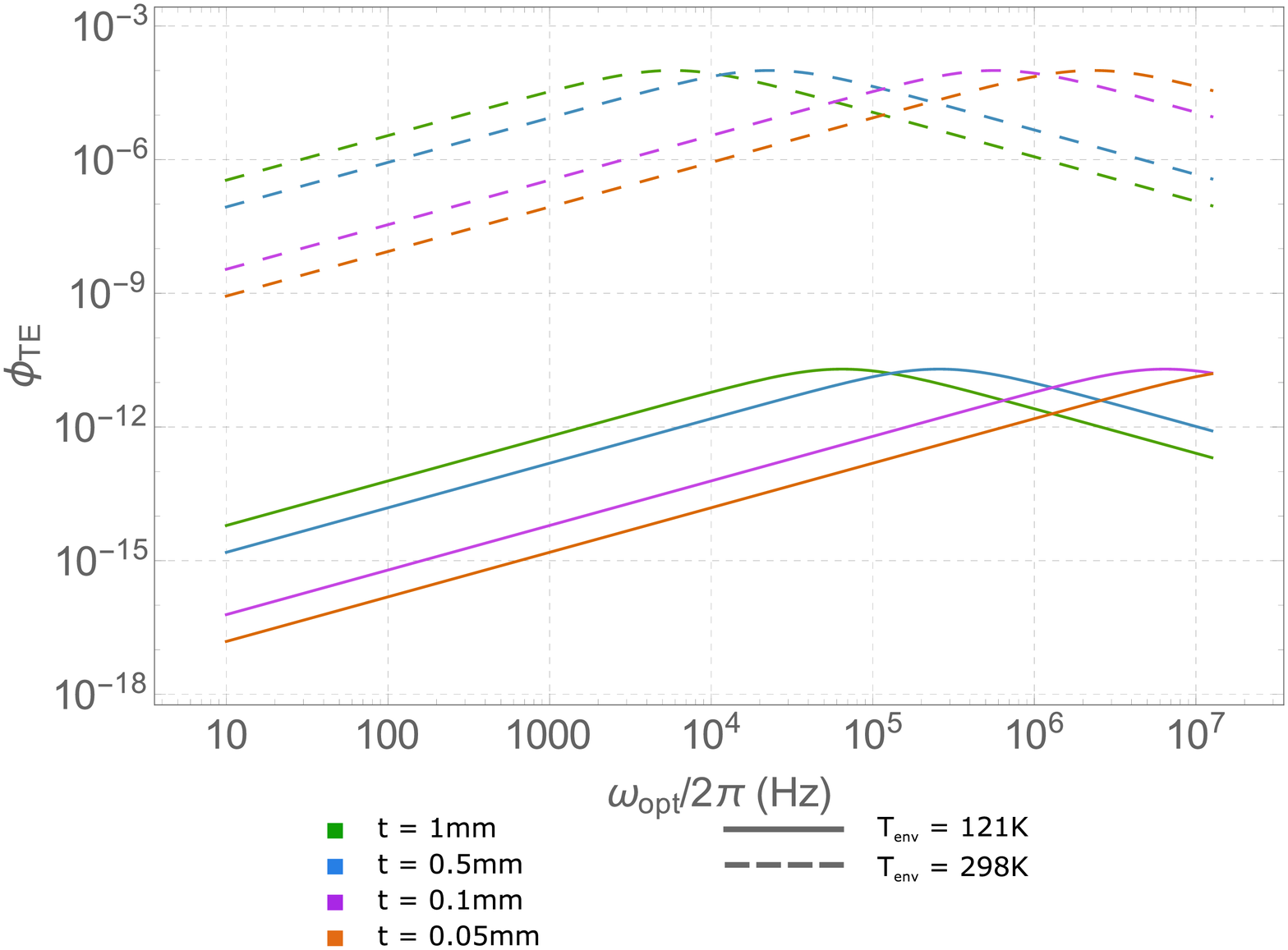}
\end{center}
\caption{Thermoelastic loss curves for various thicknesses of crystalline silicon resonators. Loss is plotted against optical spring frequency. \label{Materials2}}
\end{figure}

Flexure of the resonator causes thermoelastic loss. This loss potentially limits the \(Q_{mirror}\) of the resonator and thus the final Q-factor that can be achieved from optical dilution. It is necessary to check that the thermoelastic damping for a particular design does not exceed the intrinsic loss of the resonator. In our case, we do not wish \(Q_{mirror}\) to drop below \(10^6\), the Q-factor for crystalline mirror coatings.

To estimate the thermoelastic effects, we can use Zener’s equation from his work on flexural heat flow \cite{ZenerThermo, ZenerThermo2, LifshitzThermo}:

\begin{equation}\label{Zener}
\phi_{TE} = \frac{Y \alpha^2 T_{env}}{C_p}\times \frac{\omega_{opt}\tau}{1+\omega_{opt}^2 \tau^2}
\end{equation}

where \(\alpha\) is the coefficient of thermal expansion, \(T_{env}\) is the environmental temperature, \(C_p\) is the specific heat at constant pressure (units \(J/(m^3 K)\)) and \(\tau\) is the thermal relaxation time constant given by:

\begin{equation}\label{ThermalRelaxation}
\tau = \frac{t^2 C_p}{\pi^2 k}
\end{equation}

with \(k\) being the thermal conductivity.

We consider an assortment of materials for construction of the resonator, with their thermoelastic loss contributions shown in Figure \ref{Materials}. Throughout this paper we have used crystalline silicon due to its high intrinsic bulk quality factor. However, it has a relatively high thermal conductivity, which reduces the characteristic thermoelastic timescale. Other materials have improved thermoelastic loss at room temperature in our frequencies of interest 10 kHz to 1 MHz. Fused silica and amorphous silicon have a lower intrinsic quality factor than crystalline materials, but also have much lower thermal conductivity than crystalline silicon and end with much less thermoelastic loss than silicon at the higher frequencies.

Crystalline silicon, while relatively poor at room temperature, has an advantage in that its coefficient of thermal expansion disappears at  approximately 121 K \cite{OkadaSiliconExpansion, SwensonThermalSi, MiddlemannThermalSi}. However, the change in the coefficient is significant within one Kelvin of the zero point, so the temperature must be controlled to within a one Kelvin range. The effect of precise temperature control is shown in Figures \ref{Materials} and \ref{Materials2}. The highest thermal coefficient within the 1 Kelvin range is approximately 3 times higher than that for the 0.1 K range. The effect of thickness is seen in Figure \ref{Materials2}. Decreasing the thickness of the resonator shifts the peak of the thermoelastic loss curve towards higher frequencies.

From the calculations, it is seen that, for thicknesses lower than 0.5 mm, the thermoelastic loss of the crystalline silicon resonator at room temperature exceeds \(10^{-6}\) for optical spring frequencies of \(100\) kHz to \(1\) MHz, which is undesirable. Amorphous SiO\(_2\) has much less thermal loss and satisfies our requirements given resonator thickness approximately 0.1 mm, however, its lesser material Q-factor compared to room temperature crystalline silicon may be of concern. Crystalline silicon cooled to the zero point of thermal expansion is the best material to use at the required optical spring frequencies. Temperature control to within 1 Kelvin of the zero point is sufficient for reducing thermoelastic losses to several orders of magnitude below that of the reflective coating acoustic loss.

The recoil loss due to the coupling to the support structure could be estimated using a simple 2 stage spring-mass system. The Q-factor of a coupled pendulum has the following relation with the original Q-factor of the pendulum \(Q_p\) and supporting structure \(Q_{sup}\) \cite{SaulsonQ}:

\begin{equation}
Q_{p, recoil}^{-1} = Q_{p}^{-1} + Q_{sup}^{-1} \mu \frac{\omega_{sup} \omega_p^3}{(\omega_{sup}^2 - \omega_p^2)^2}
\end{equation}

With \(\mu = m_p / m_{sup}\). Quantities with subscript \(p\) correspond to the properties of the optical spring enhanced resonator. The loss added to the pendulum Q-factor becomes negligible if the mass ratio is very low and the frequency difference \(\omega_{sup}^2 - \omega_p^2\) is very high.  One strategy of reducing vibration losses is to rigidly attach the support structure to a vibrationally isolated breadboard with mass 10--100 kg. For a microgram resonator with frequency \(2\pi \times 100\) kHz and 10 kg breadboard suspended by a pendulum of frequency \(2\pi \times 10\) Hz, the coefficient of \(Q_{sup}^{-1}\) is approximately \(10^{-14}\). Thus, the recoil loss can be made insignificant with the correct choice of mass ratio and frequency difference.

\begin{table*}\begin{center}\begin{ruledtabular}
\begin{tabular}{|l|p{6cm}|p{6cm}|}
\textbf{Loss Mechanism}&\textbf{Description}&\textbf{Mitigation} \\ \hline
Recoil loss & Recoil imparted onto the pendulum by coupling to the support structure & Optical spring beams must align close to centre of percussion. Decrease mass ratio of pendulum versus support. \\ \hline
Torsion and tilt & Unwanted modes of pendulum motion & Reduce the coupling of these modes by careful alignment of the beam. \\ \hline
Gas pressure loss & Damping of motion by gas molecules & UHV ion pumping to sufficiently low pressure, depending on resonator dimension and desired optical spring frequency. \\ \hline
Suspension loss & Suspension internal modes that couple to centre mass (CM) motion & Minimise membrane thickness and length to keep unwanted resonances far above 200 kHz, maximise mass ratio between mirror and membrane. Use centre of percussion tuning to minimise the coupling factor. \\ \hline
Thermoelastic damping & Nonuniform temperature distribution from vibration flexure causes heat flow and internal friction \cite{ZenerThermo, ZenerThermo2, LifshitzThermo} & Reduce \(\alpha(T)\) of crystalline silicon through cooling or use amorphous SiO\(_2\) for the suspended mass  \\ \hline
Diffraction loss & Diffractive loss from the edges of the mirror & Keep mirror radius 3 times larger than \(1/e^2\) beam power radius on the mirror, use tapered edges on mirror \cite{ChangTrapping} \\ \hline
Acceleration loss & Acceleration causes mechanical loss through resonator deformation &Small resonator dimensions to maintain CM motion and minimise deformation, minimise coating loss \\ \hline
Charge and magnetic coupling & Charge buildup on the resonator & Keep conductors at large distance. Possible UV charge neutralisation techniques. \\
\end{tabular}\end{ruledtabular}\end{center}
\caption{Loss mechanisms for the suspended resonator}
\label{Losses}
\end{table*}

Other mechanisms of loss are summarised in Table \ref{Losses}. Electromagnetic coupling can be reduced by appropriate shielding, conductor placement and charge neutralisation techniques. 
Recoil loss can be minimised by centre of percussion tuning, but given that the mass ratio between the resonator and support can be made such that recoil loss is sufficiently reduced, optical beam alignment can then be used to reduce torsion and tilt of the pendulum. Diffraction loss increases when the resonator size is comparable to the beam size. Careful design the DEMS cavity will make the beam size on the resonator small enough to ensure the cavity finesse no reduction due to the diffraction loss. The beam size on the mirror should be such that the \(1/e^2\) beam power diameter is less than one third of the size of the mirror, which gives less than 1ppm diffractive power loss.

\section{Fabrication of a resonator with nanoscale silicon nitride suspension}\label{SuspensionFabrication}

\begin{figure}
\begin{center}
\includegraphics[width = 0.4\textwidth]{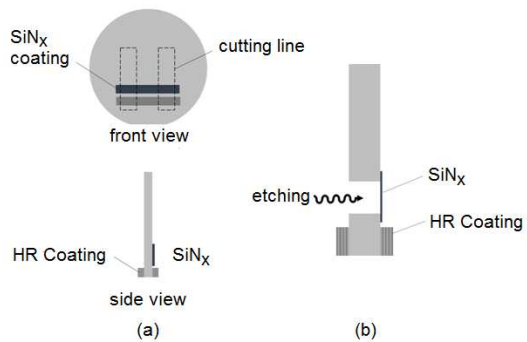}
\end{center}
\caption{Major steps of the fabrication process for the resonator (a) HR coating and SiN\(_x\) applied to a silicon wafer (b) Etching of bulk silicon to form the resonator.\label{Wafer}}
\end{figure}

\begin{figure}
\begin{center}
\includegraphics[width = 0.4\textwidth]{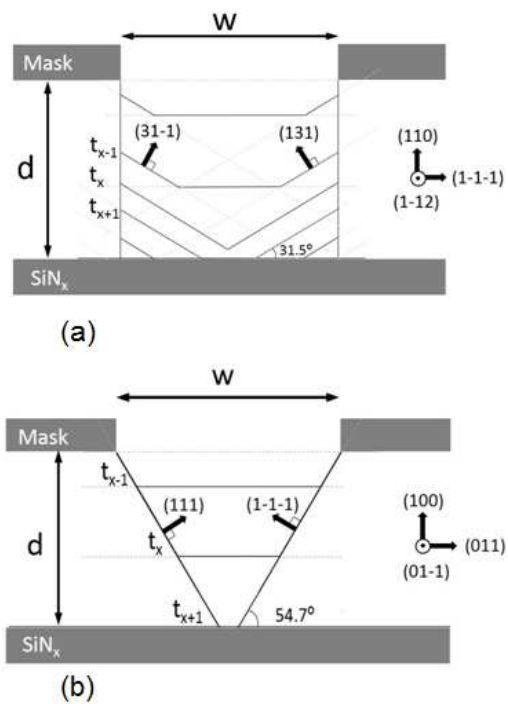}
\end{center}
\caption{Close-up views and time evolution for etching on (a) (110)oriented wafer and (b) (100) wafer.\label{Faces}}
\end{figure}

We have described above the mechanics of a resonator with extremely low suspension to the thermal reservoir. The suspension membrane thickness is the primary source of the mechanical stiffness of the resonator and thus must be minimised. The fabrication of this suspension must be done such that the weight can be supported and the geometry of the configuration does not undergo unnecessary stresses, such as twisting or cracking.

The cat-flap resonator requires a small but macroscopic mass to be suspended with a nanoscale suspension element in the form of a very thin membrane or nanowires. There are several potential methods of suspension such as a graphene or silicon nitride membrane, carbon nanotubes or SiN strips. Here we demonstrate a possible approach based on construction of a silicon mirror suspended by an SiN membrane. Figure \ref{Wafer} shows the general process for constructing a membrane suspended resonator.


Wet or dry etching can be used to remove silicon and create the membrane suspension. One method of dry etching is reactive ion etching (RIE), where a reactive plasma mixture, such as SF\(_6\) and O\(_2\), is used to selectively remove silicon - the rate of silicon to SiN\(_x\) is 10:1 \cite{ChaoRate}. The rate of reaction means that the end-point of the etching must be monitored, otherwise defects on the SiN membrane will be produced. This process results in imperfections on the silicon faces adjacent to the membrane such as under-cutting, but this can be mitigated with a C\(_4\)F\(_8\) passivation layer \cite{ChaoEtch1, ChaoEtch2, ChaoEtch3}.

Wet etching in KOH allows for better selectivity of material removal, since the rate of etching of SiN is negligible \cite{ChaoRate}, however, the rate of silicon removal depends upon the crystallographic properties \cite{Chao6, Chao7}. The \{111\} family has a much smaller etching rate than other faces, and so should be used for the side-walls of the ``trench''. The most easily obtainable commercial wafers are oriented (110) or (100) on the large face, and thus we design fabrication processes for these two. The orientations of the faces are shown in Figure \ref{Faces}.

For the (110) process, the \{311\} faces are removed more slowly than the others, apart from \{111\} \cite{Chao6}. Thus, the process is dependent upon the \{311\} faces. As shown in Figure \ref{Faces} (a), the silicon is removed starting from the centre and working to the sides of the trench, until the KOH reaches the \{111\} face defined by the mask. This forms a well defined rectangular spacing. Figure \ref{Faces} (b) shows the reaction profile for the (100) wafer. In this orientation, the \{111\} faces are diagonal and converging, which means that if the thickness of the wafer is too high, the \{111\} diagonals will meet and the reaction will not proceed any further. The width of the spacing can be found if the nominal spacing width and wafer thickness are known. The (100) process gives a trapezoidal rather than rectangular spacing.

For KOH etching, the width of the strip needs to be cut wider than the design value to account for unavoidable etching of the edges. The cutting must yield an extremely smooth and straight edge so that the extra removal will be uniform. Jagged edges, having multiple crystallographic faces, will become rougher if KOH is applied. If this happens, the method of cleavage must be applied for cutting.

The materials used for the resonator must be able to withstand photolithography. HR coatings must be protected throughout the whole process and the protective material must be removed without residue or damage. It could also be possible to deposit the HR coating after the etch but this puts limitations on the membrane flexure design.

\begin{figure*}
\begin{center}
\includegraphics[width=0.9\textwidth]{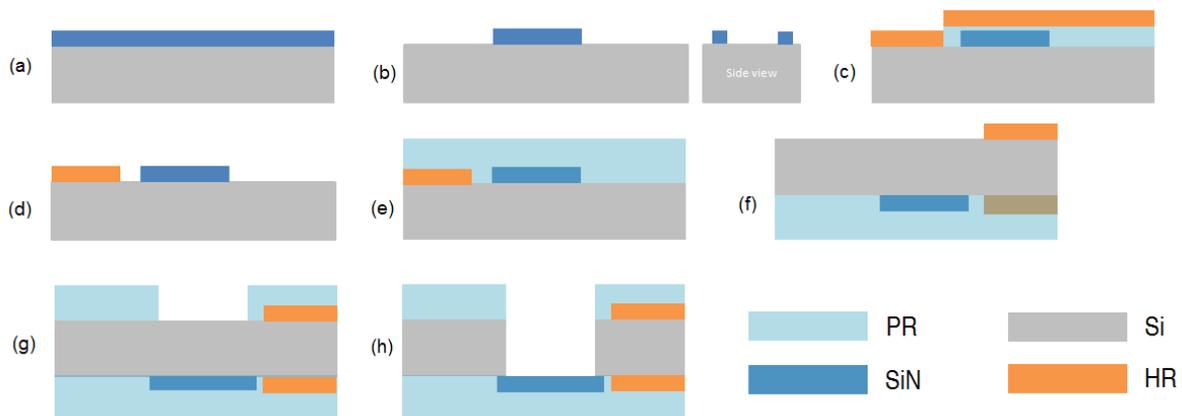}
\end{center}
\caption{The fabrication process for silicon nitride nanowire suspension. Note that the diagrams show the resonator in the horizontal orientation rather than vertical. \textbf{(a)}: SiN deposition \textbf{(b)}: Nanowire patterning \textbf{(c)}: Depositing of lift-off layer and 
HR coating \textbf{(d)}: Photoresist lift-off \textbf{(e)}: Top layer protection \textbf{(f)}: Backside HR patterning \textbf{(g)}: Backside lithography - defining mask for silicon etching \textbf{(h)}: silicon etching to form the pendulum structure. Note that this figure does not show undercutting as a result of the etching process. \label{Nanowires}} 

\end{figure*}

Silicon nitride nanowires can be used to suspend smaller resonators. The entire process is shown in Figure \ref{Nanowires}. SiN is deposited onto a (100) silicon wafer using LPCVD, then patterned into nanowires. The patterning is performed using a mask with RIE, and the HR coating is patterned with the lift-off process. The entire front side is then covered with a protective photoresist, and the HR coating process is repeated for the opposite mirror face. Once both sides have undergone HR coating, the photoresist mask is then prepared for the etching of the silicon to create the pendulum structure. Due to the etching process, undercuts will be present as in the (100) wafer method for the silicon nitride membrane. However, these are relatively small changes in geometry, and will not have a significant effect on the quality factor of the mirror, as demonstrated by gravitational wave test masses with bonded ears \cite{GEOEars}.

Since the device must be transported, we must ensure that it does not break in transit. It is possible to leave the photoresist layers on the front and back sides of the resonator after the silicon etching step, and transport the device horizontally as shown in Figure \ref{Nanowires}. To prepare the resonator for experiments, it can be tilted into the correct orientation and then immersed in nitric acid to remove the protective photoresist layer.

SiN\(_x\) tensile strength is dependent upon the fabrication process. Plasma enhanced chemical vapour deposition (PECVD) fabrication gives a strength of 390 MPa for silicon-rich composition and 420 MPa for nitrogen-rich composition \cite{Chao8, Chao9}. Low pressure CVD (LPCVD) fabrication reportedly allows strengths as high as 5--8 GPa \cite{Chao9, Chao10}. The estimated tensile stress for a 1 mm wide, 10 nm thick membrane supporting a 2 mg mass is 2 MPa, two orders of magnitude below the tensile strength for PECVD silicon nitride. For two 50 \(\times\) 50 nm\(^2\) LPCVD nanowires, the largest mass that can be supported, given a safety factor of 5, is 0.5 mg, approximately the mass of a 1 \(\times\) 1 \(\times\) 0.25 mm resonator.

\section{Discussion}

Equation \ref{Dilution} shows that, disregarding the losses discussed in Section \ref{MechN}, the optical spring trap increases the quality factor proportional to the square of the optical spring frequency. This proportionality operates under the assumption that the mechanical damping is structural rather than viscous. Under the viscous damping regime, the quality factor from an optical spring scales linearly with \(\omega_{opt}\) \cite{KorthSpring}. Ni, \textit{et al.} \cite{NiTrap} have used optical trapping to increase the Q-factor of a pendulum resonator 50-fold, which was greater than their optical trap frequency ratio \(\omega_{opt}/\omega_m \sim 23\), but less than the square of this frequency ratio. However, this experiment also featured significant suspension losses, so the proportion of viscous to structural damping was unclear. Further experimentation with cat-flap type resonators may be necessary to determine the extent of viscous versus structural damping. 

The description of acceleration loss in Section \ref{AccelLoss} incorporates a simple one-dimensional analysis. Within the resonator only longitudinal modes were considered. However, in practice, an optical spring may interact with 2 or 3-dimensional modes such as bending and twisting modes. 3-dimensional analysis of deformations in a resonator would allow for a more accurate estimate for the effects of acceleration loss upon the optical spring \(Q_f\).

Section \ref{AccelLoss} describes the limitation of optical dilution by acceleration loss. We estimated attainable quality factors for various size resonators. Apart from optimisation, another solution to this problem lies in the construction of the dielectric quarter wave stack. Many of these stacks are designed to reflect a high percentage of light from the first few layers, with the additional layers improving the reflectivity by a small amount. This causes a large radiation pressure on the surface of the mirror. Instead, we propose using a stack which reflects equal amplitudes of light at each layer. The reflectivity of the first surface should be low, with subsequent surfaces increasing in reflectivity. This will then result in a more uniformly distributed radiation pressure with less acceleration losses.

A consequence of this construction is the reduction of thermal noise. Deformation and thermal noise are intrinsically linked. A mechanical fluctuation produces thermal noise, and vice versa. Creating a dielectric stack which reduces the compressive deformation of the laser can also reduce the thermal noise. It should be noted that these effects have been estimated in one dimension, so three dimensional effects will limit the reduction of the compressive loss and thermal noise.

Another effect that is not considered in detail is gravitational dilution. In a manner similar to optical dilution, the coupling of motion to a lossless field increases the quality factor higher than the material Q-factor \cite{MitrofanovDilution, CagnoliDilution}. For a pendulum, the motion is coupled to the gravitational field, with high Q-factors possible for thinner suspensions. The presence of gravitational dilution will increase the mechanical Q-factor \(Q_m\) used throughout this paper for the  centre-of-mass pendulum motion with no applied laser.

In Section \ref{Comparisons} we discuss the effect of suspension violin string modes, using parameters estimated from carbon nanotubes. However, these parameters are prone to uncertainty given the variations in their construction. The Young's Modulus for single-walled carbon nanotubes is approximately 1.3 TPa, with measurements from the vibrational characteristics producing a derived Young's modulus of \(1.3^{+0.6}_{-0.4}\) TPa \cite{KrishnanNanotubes}. Compressive loading experiments derive a Young's modulus of 2.8-3.6 TPa \cite{BuckyCentral}. The thickness used in the derivations in Section \ref{Comparisons}  is usually taken as 0.34nm, the interlayer spacing of graphene, however, some computational methods derive the thickness as being 0.06-0.07 nm, using the well defined stiffness and simulated Young's modulus \cite{HuangNanotubes}. For the purposes of this analysis we used values derived from vibrational amplitude experiments. Since these are subject to uncertainties of up to 50\%, combined with assumptions regarding tube properties, such as isotropic elasticity, our calculations remain as an order of magnitude estimate.

\section{Conclusion}

We have presented an analysis for an optical dilution scheme that is predicted to enable extremely high quality factor millimetre to sub-millimetre scale resonators in the frequency range \(10^3\) to \(10^5\) Hz. Two new concepts have been introduced: the Double End-Mirror Sloshing (DEMS) cavity and a miniature pendulum called a cat-flap resonator. In addition, we have mentioned a mirror coating scheme that can suppress acceleration loss and reduce thermal noise. This will be the subject of a future paper.

The DEMS cavity has been shown to allow cancellation of quantum noise by correct choice of sloshing frequency and detuning. Stable optical trapping and dilution of thermal noise was shown to be possible for the DEMS cavity. The quantum radiation pressure noise can be reduced by a factor of \(\gamma_e/\gamma_f\).

Creation of successful thermal noise free optomechanics will require careful optimisation of resonator/suspension design and experimental parameters. We analysed the approximate magnitude of acceleration losses which are a limiting factor for all trapped mechanical resonators. The acceleration loss limits the optical spring stiffened Q-factor to a maximum attainable \(Q_f\) of \(10^{14}\) at frequencies of 1-10 kHz, for resonators of dimension \(d=0.1\) mm or lower, under UHV conditions of \(10^{-11}\) Torr. Resonators with dimension \(d<0.1\) mm are also able to achieve a maximum Q-factor of \(10^{11}\) at \(\omega_{opt}=100\) kHz and pressure of \(10^{-8}\) Torr. Achievement of such performance requires use of low acoustic loss mirror materials such as crystalline GaAs mirror coatings \cite{ColeCrystalline}. Thermoelastic noise could be a problem for crystalline silicon resonators unless they are cooled to 120 K where thermoelastic noise falls to zero. There is currently insufficient information about loss of nanowires and membranes to be able to make precise predictions of the final quality factor.

Fabrication of silicon nitride suspension has been shown possible for the dimensions described in this paper. We discussed the advantages of carbon nanotube suspension, however, many technical problems need to be solved to be able to use them. Another interesting suspension material would be graphene membranes. Efforts are underway to develop SiN-suspended resonators described in this paper by authors SC and HS.

If ultra-high Q-factor mechanical resonators can be successfully implemented using the methods discussed here, then it should be relatively straightforward to create white light cavities that will enable substantial enhancement in the sensitivity of gravitational wave detectors.

{\bf Acknowledgements}

We wish to thank the Optics Working Group of LIGO Scientific Collaboration for advice, and Dr Giles Hammond for useful discussions. University of Western Australia research was supported by the Australian Research Council (Grants No. DP120104676 and No. DP120100898). Authors Pan and Chao were supported by the Ministry of Science and Technologies of Taiwan, Republic of China (MOST 103-2221-E-007-064-MY3). Author Mitrofanov was supported by the  Russian Foundation for Basic Research (14-02-00399). Author Sadeghian was supported by the Enabling Technology Program (ETP) Optomechatronics and Early Research Program (ERP) 3D Nanomanufacturing at TNO.

\bibliography{physRefs}
\end{document}